\documentstyle[preprint,floats,pra,aps,epsfig]{revtex} 
\tightenlines
\begin{document}

\draft

\title{\LARGE \bf Finite Difference Distributions for Ginibre Ensemble} 

\author{Maciej M. Duras 
\thanks{Electronic address: mduras @ riad.usk.pk.edu.pl}} 
\address{Institute of Physics, Cracow University of Technology, 
ulica Podchor\c{a}\.zych 1, PL-30-084 Cracow, Poland} 

\date{1st March, 2000}
  
\maketitle 

\begin{center}
{\em J. Opt. B: Quantum Semiclass. Opt.}  {\bf 2}, 287-291 (2000). 
\end{center}

\begin{abstract}
The Ginibre ensemble of complex random matrices is studied.
The complex valued random variable
of second difference of complex energy levels is defined.
For the N=3 dimensional ensemble are calculated distributions 
of second difference, of real and imaginary parts of second difference,
as well as of its radius and of its argument (angle).
For the generic N-dimensional Ginibre ensemble
an exact analytical formula for second difference's
distribution is derived.
The comparison with real valued random variable 
of second difference of adjacent real valued energy levels
for Gaussian orthogonal, unitary, and symplectic, 
ensemble of random matrices
as well as for Poisson ensemble is provided.
\end{abstract} 

\pacs{05.30.Ch, 05.30.-a, 02.50.-r, 05.10.-a}

\section{Introduction}
\label{sec-introduction}
Random Matrix theory assumes that the Hamiltonian operator $H$
of a generic quantum system is unknown and unknowable
\cite{Haake 1990,Guhr 1998,Mehta 1990 0}.
The matrix elements $H_{ij}$ of Hamiltonian in given basis
of Hilbert space are random variables.
Their distributions are given by appropriate formulae
depending on studied Random Matrix ensemble
\cite{Haake 1990,Guhr 1998,Mehta 1990 0}.
Thy symmetry properties of $H$ which is hermitean
lead us to Gaussian ensembles of random matrices:
orthogonal GOE, unitary GUE, symplectic GSE,
as well as to circular ensembles: orthogonal COE,
unitary CUE, and symplectic CSE.
The energies $E_{i}$ of quantum systems calculated from
diagonalization of Hamiltonian matrix $H_{ij}$
are random variables with appropriate distributions
and they exhibit generic classes of level repulsion.
It was Wigner who firstly discovered level repulsion
phenomenon \cite{Haake 1990,Guhr 1998,Mehta 1990 0}.
The applications of Random Matrix theory are very broad:
nuclear physics (slow neutron resonances, highly excited complex nuclei),
condensed phase physics (fine metallic particles,  
random Ising model [spin glasses]),
quantum chaos (quantum billiards, quantum dots), 
disordered mesoscopic systems (transport phenomena). 
J. Ginibre studied very general case of random Hamiltonians.
He dropped the assumption of hermiticity of Hamiltonians
and he considered generic complex valued matrices
\cite{Haake 1990,Guhr 1998,Ginibre 1965,Mehta 1990 1}.
Thus, $H$'s belong to general linear Lie group GL(N, {\bf C}),
where N is dimension and {\bf C} is complex numbers field.
Therefore, the energies $Z_{i}$ of quantum system 
ascribed to Ginibre ensemble are complex valued.
This is an extension of Gaussian or circular ensembles.
J. Ginibre postulated the following
joint probability density function 
of random vector of complex eigenvalues $Z_{1}, ..., Z_{N}$
for $N \times N$ Hamiltonian matrices 
\cite{Haake 1990,Guhr 1998,Ginibre 1965,Mehta 1990 1}:
\begin{equation}
P(z_{1}, ..., z_{N})=
\prod _{j=1}^{N} \frac{1}{\pi \cdot j!} \cdot
\prod _{i<j}^{N} \vert z_{i} - z_{j} \vert^{2} \cdot
\exp (- \sum _{j=1}^{N} \vert z_{j}\vert^{2}).
\label{Ginibre-joint-pdf-eigenvalues}
\end{equation}
We emphasise that 
$Z_{i}$'s are {\it complex valued} random variables, 
and $z_{i}$'s are {\it complex} sample points
($z_{i} \in {\bf C}$).
 
One must emphasise here Wigner and Dyson's electrostatic analogy.
A Coulomb gas of N unit charges moving on complex plane (Gauss' plane)
{\bf C} is considered. The vectors of positions
of charges are $z_{i}$'s and potential energy of the system is:
\begin{equation}
U(z_{1}, ...,z_{N})=
- \sum_{i<j} \ln \vert z_{i} - z_{j} \vert
+ \frac{1}{2} \sum_{i} \vert z_{i}^{2} \vert. 
\label{Coulomb-potential-energy}
\end{equation}
If gas is in thermodynamical equilibrium at temperature
$T= \frac{1}{2 k_{B}}$ 
($\beta= \frac{1}{k_{B}T}=2$, $k_{B}$ is Boltzmann's constant),
then probability density function of vectors of positions is 
$P(z_{1}, ..., z_{N})$ Eq. (\ref{Ginibre-joint-pdf-eigenvalues}).
Thus, complex energies of quantum system 
and vectors of positions of charges of Coulomb gas are analogous
to each other.
In view of above analogy one can consider the complex spacings
$\Delta^{1} Z_{i}$ of complex
energies of quantum system:
\begin{equation}
\Delta^{1} Z_{i}=Z_{i+1}-Z_{i}, i=1, ..., (N-1),
\label{first-diff-def}
\end{equation}
as vectors of relative positions of electric charges of
Coulomb gas.
For Ginibre ensemble it were calculated 
the distributions of {\it real valued} absolute values of spacings
of nearest neighbour {\it ordered} energies. 
We complement this by introduction of complex valued
second differences $\Delta^{2} Z_{i}$ of complex energies:
\begin{equation}
\Delta ^{2} Z_{i}=Z_{i+2} - 2Z_{i+1} + Z_{i}, i=1, ..., (N-2).
\label{Ginibre-second-difference-def}
\end{equation}
The second differences are three energy level magnitudes
that enhance our knowledge of quantum systems
with complex energies. Moreover, $\Delta ^{2} Z_{i}$'s
can be regarded as vectors of relative positions of vectors
of relative positions of electric charges.
One can observe movement of electric charges is Cartesian
frame of references or in polar one.
Since the two-dimensional vectors in Cartesian frame of reference
have their projections on
co-ordinate axes, therefore the real and imaginary parts of
$\Delta ^{2} Z_{i}$'s, namely ${\rm Re} \Delta ^{2} Z_{i}$,
${\rm Im} \Delta ^{2} Z_{i}$, can be interpreted 
as projections of second differences 
on abscissa and ordinate axes, respectively.
The radii $\vert \Delta ^{2} Z_{i} \vert$,
and arguments (angles) ${\rm Arg} \Delta ^{2} Z_{i}$ of second differences
have interpretations of polar co-ordinates of $\Delta ^{2} Z_{i}$ vectors.
$\Delta ^{2} Z_{i}$'s are analogous to real valued second differences:
\begin{equation}
\Delta^{2} E_{i}=E_{i+2}-2E_{i+1}+E_{i}, i=1, ..., (N-2),
\label{second-diff-def}
\end{equation}
of adjacent ordered increasingly real valued energies $E_{i}$
defined for
GOE, GUE, GSE, and Poisson ensemble PE
(where Poisson ensemble is composed of uncorrelated
randomly distributed energies)
\cite{Duras 1996 PRE,Duras 1996 thesis,Duras 1999 Phys,Duras 1999 Nap}.
We will calculate the distributions of
$\Delta ^{2} Z_{i}$,
${\rm Re} \Delta ^{2} Z_{i}$, 
${\rm Im} \Delta ^{2} Z_{i}$,
$\vert \Delta ^{2} Z_{i} \vert$, 
${\rm Arg} \Delta ^{2} Z_{i}$.
Finally, we will compare these results with
second difference distributions
for Gaussian ensembles, and Poisson ensemble
\cite{Haake 1990,Mehta 1990 2,Reichl 1992,Porter 1965 Gaussian 1,Wigner 1957,Wigner
1956,Porter Rosenzweig 1960 Suomalaisen}.

\section{Second Difference Distributions}
\label{sec-second-difference-pdf}
We use formula (\ref{Ginibre-joint-pdf-eigenvalues})
with $N=3$ and define the following complex valued random vector $(Y_{1}, Y_{2}, Y_{3})$
and real $A_{j}$ and imaginary $B_{j}$ parts:
\begin{equation}
Y_{1}=\Delta ^{2} Z_{1}, Y_{2}=Z_{2} - Z_{3}, Y_{3}=Z_{3}, 
Y_{j}=(A_{j}, B_{j}), A_{j}= {\rm Re} Y_{j}, B_{j}= {\rm Im} Y_{j}, j=1, ...,3.
\label{Ginibre-Y-def}
\end{equation}
The change of the variable formula gives us the result for 
joint probability density function of random vector $(Y_{1}, Y_{2}, Y_{3})$
\cite{Bickel 1977}:
\begin{eqnarray}
& & f_{(Y_{1}, Y_{2}, Y_{3})}(y_{1}, y_{2}, y_{3})=
f_{(A_{1}, B_{1}, A_{2}, B_{2}, A_{3}, B_{3})}(a_{1}, b_{1}, a_{2}, b_{2}, a_{3}, b_{3})=
\label{Ginibre-joint-pdf-Y}
\\
& & \frac{1}{12 \pi^{3}} \cdot
[(a_{1}+ a_{2})^{2} + (b_{1}+ b_{2})^{2}] \cdot
[a_{2}^{2} + b_{2}^{2}] \cdot
[(a_{1}+ 2a_{2})^{2} + (b_{1}+ 2b_{2})^{2}] \cdot
\nonumber 
\\ 
& & \cdot [ \exp (- (a_{1}+ 2a_{2} + a_{3})^{2} - (b_{1}+ 2b_{2} + b_{3})^{2}
- (a_{2} + a_{3})^{2} - (b_{2} + b_{3})^{2}- a_{3}^{2} - b_{3}^{2})],
\nonumber
\end{eqnarray}
where $y_{j}=(a_{j}, b_{j}) \in {\bf C}$ are complex random sample points.
[In order to obtain Eq. (\ref{Ginibre-joint-pdf-Y}) 
we used the complex valued linear map
\begin{equation}
(Y_{1}, Y_{2}, Y_{3})= \Xi (Z_{1}, Z_{2}, Z_{3}),
\Xi = 
\left[
\begin{tabular}{ccc}
1 & -2 & 1 \\
0 & 1 & -1 \\
0 & 0 & 1 \\
\end{tabular}
\right],
\label{Ginibre-linear-map-def}
\end{equation} 
where Jacobian of the inverse map $\Xi^{-1}$ 
is equal to unity ${\rm Jac} (\Xi^{-1}) = 1$].
We integrate out $Y_{3}$:
\begin{equation}
f_{(Y_{1}, Y_{2})}(y_{1}, y_{2})=
\int _{{\bf C}} f_{(Y_{1}, Y_{2}, Y_{3})}(y_{1}, y_{2}, y_{3}) dy_{3},
\label{Ginibre-joint-pdf-Y1Y2-def}
\end{equation}
and we obtain following marginal probability density function:
\begin{eqnarray}
& & f_{(Y_{1}, Y_{2})}(y_{1}, y_{2})=
f_{(A_{1}, B_{1}, A_{2}, B_{2})}(a_{1}, b_{1}, a_{2}, b_{2})= 
\label{Ginibre-joint-pdf-Y1Y2-result}
\\
& & 
\frac{1}{36 \pi^{2}}
[(a_{1}+a_{2})^{2}+(b_{1}+b_{2})^{2}]
[a_{2}^{2}+b_{2}^{2}]
[(a_{1}+2a_{2})^{2}+(b_{1}+2b_{2})^{2}] \cdot
\nonumber
\\
& &
\cdot \exp [ -\frac{2}{3}a_{1}^{2} -2a_{1}a_{2} -2a_{2}^{2}
-\frac{2}{3}b_{1}^{2} -2b_{1}b_{2} -2b_{2}^{2} ].
\nonumber
\end{eqnarray}
Now we calculate the marginal probability density function of
second difference:
\begin{equation}
f_{Y_{1}}(y_{1})=f_{(A_{1}, B_{1})}(a_{1}, b_{1})=
\int _{{\bf C}} f_{(Y_{1}, Y_{2})}(y_{1}, y_{2}) dy_{2}
=\frac{1}{576 \pi} [ (a_{1}^{2} + b_{1}^{2})^{2} + 24]
\cdot \exp (- \frac{1}{6} (a_{1}^{2}+b_{1}^{2})).
\label{Ginibre-marginal-pdf-Y1-def}
\end{equation}

Now we derive the distributions of real part $A_{1}$
and of imaginary part $B_{1}$ of second difference:
\begin{equation}
f_{A_{1}}(a_{1})=
\int _{{\bf R}} f_{(A_{1}, B_{1})}(a_{1}, b_{1}) db_{1}
=\frac{\sqrt{6}}{576 \sqrt{\pi}} (a_{1}^{4}+6a_{1}^{2}+ 51)
\cdot \exp (- \frac{1}{6} a_{1}^{2}),
\label{Ginibre-marginal-pdf-Y1Re-def}
\end{equation}
\begin{equation}
f_{B_{1}}(b_{1})=
\int _{{\bf R}} f_{(A_{1}, B_{1})}(a_{1}, b_{1}) da_{1}
=\frac{\sqrt{6}}{576 \sqrt{\pi}} (b_{1}^{4}+6b_{1}^{2}+ 51)
\cdot \exp (- \frac{1}{6} b_{1}^{2}),
\label{Ginibre-marginal-pdf-Y1Im-def}
\end{equation}
where ${\bf R}$ is field of real numbers.

We transform complex valued random variable of second difference $Y_{1}$ 
to polar co-ordinates' variables $R_{1}, \Phi_{1}$:
\begin{equation}
R_{1} = \vert Y_{1} \vert, \Phi_{1}= {\rm Arg} Y_{1},
\label{Ginibre-polar-second-diff-def}
\end{equation}
and we obtain by standard method the following probability density function
of random vector $(R_{1}, \Phi_{1})$:
\begin{equation}
f_{(R_{1}, \Phi_{1})}(r_{1}, \phi_{1})=
\Theta(r_{1}) \frac{1}{576 \pi}r_{1}(r_{1}^{4}+24) 
\cdot \exp(- \frac{1}{6} r_{1}^{2})
\label{Ginibre-polar-second-diff-R1Phi1}
\end{equation}
(the Jacobian of transformation is $r_{1}$,
$\Theta$ is Heaviside (step) function \cite{Bickel 1977}).
It follows that:
\begin{equation}
f_{R_{1}}(r_{1})=
\Theta(r_{1}) \frac{1}{288}r_{1}(r_{1}^{4}+24) \cdot \exp(- \frac{1}{6} r_{1}^{2}),
f_{\Phi_{1}}(\phi_{1})= \frac{1}{2 \pi}, \phi_{1} \in [0, 2 \pi].
\label{Ginibre-polar-second-diff-result}
\end{equation}

\section{N-dimensional Ginibre ensemble}
\label{sect-N-dimensional}

The case of generic N-dimensional Ginibre ensemble
is of special physical interest (N $\geq 3$).
We will calculate the distribution of
second difference for the ensemble.
One commences with n-level correlation function:
\begin{eqnarray}
& & P_{n}(z_{1}, ..., z_{n})=
\int_{{\bf C}^{N-n}} P(z_{1}, ..., z_{N}) dz_{n+1} ... dz_{N}=
\label{n-level-correlation-function-definition} \\
& & = \pi^{-n} \exp(- \sum_{i=1}^{n} \vert z_{i} \vert^{2})
\det D^{(n)},
\label{n-level-correlation-function-result} \\
& & D_{ij}^{(n)}=e_{N-1}(z_{i}z_{j}^{*}),
i, j=1, ...,n,
e_{N-1}(z)=\sum_{k=0}^{N-1} \frac{z^{k}}{k !},
\label{n-level-correlation-function-D-e}
\end{eqnarray}
Ref. \cite{Haake 1990}.
In order to calculate the distribution of
complex second difference $W_{1}=\Delta ^{2} Z_{1}$
for N-dimensional Ginibre ensemble one substitutes
n=3 to Eq. (\ref{n-level-correlation-function-definition})
and defines random vector $W=(W_{1}, W_{2}, W_{3}):$
\begin{equation}
(W_{1}, W_{2}, W_{3})= \Omega (Z_{1}, Z_{2}, Z_{3}),
\Omega = 
\left[
\begin{tabular}{ccc}
1 & -2 & 1 \\
0 & 1 & 0 \\
0 & 0 & 1 \\
\end{tabular}
\right].
\label{Ginibre-linear-map-def-N-dim}
\end{equation} 
The probability density function of random vector $W$ reads:
\begin{equation}
P_{3}(w_{1}, w_{2}, w_{3})=
P_{3}(\Omega^{-1}(w_{1}, w_{2}, w_{3}))=
P_{3}(w_{1}+2w_{2}-w_{3}, w_{2}, w_{3}),
\label{W-pdf}
\end{equation}
Ref. \cite{Bickel 1977}.
Hence, the distribution of second difference is:
\begin{equation}
P_{3}(w_{1})= 
\int_{{\bf C}^{2}} P_{3}(w_{1}+2w_{2}-w_{3}, w_{2}, w_{3}) dw_{2}dw_{3}.
\label{W1-pdf}
\end{equation}
We combine Eqs (\ref{n-level-correlation-function-definition}), 
(\ref{n-level-correlation-function-result}), 
(\ref{n-level-correlation-function-D-e}), 
(\ref{W-pdf}), (\ref{W1-pdf}),
and we use Laplace's expansion of determinant
$\det D^{(n)}$:
\begin{equation}
P_{3}(w_{1})= 
\pi^{-3} \sum_{{\cal{P}}}
(-1)^{{\cal{P}}}
\int_{{\bf C}^{2}}
\exp(-\sum_{i=1}^{3} \vert (\Omega^{-1} w)_{i} \vert^{2})
\prod_{k=1}^{3}
e_{N-1}[(\Omega^{-1} w)_{k} \cdot (\Omega^{-1} w)_{{\cal{P}}k}^{*}]dw_{2}dw_{3},
\label{W1-pdf-permutation}
\end{equation}
where ${\cal{P}}$ is permutation of indices $(1,2,3)$.
The only nonzero contribution to Eq. (\ref{W1-pdf-permutation})
is for identity permutation ${\cal{P}}={\rm id}=(1,2,3)$.
It results from the fact that other permutations produce factors that are
periodic functions of arguments ${\rm Arg} w_{2}, {\rm Arg} w_{3}$ 
of complex numbers $w_{2}, w_{3}$
(the integrals over $w_{2}, w_{3}$ can be transformed to polar
co-ordinates where the arguments ${\rm Arg} w_{2}, {\rm Arg} w_{3}$
are integrated over $[0, 2 \pi]$). 
Hence,
\begin{equation}
P_{3}(w_{1})= 
\pi^{-3} 
\int_{{\bf C}^{2}}
\exp(-\sum_{i=1}^{3} \vert (\Omega^{-1} w)_{i} \vert^{2})
\prod_{k=1}^{3}
e_{N-1}[\vert (\Omega^{-1} w)_{k} \vert^{2}]dw_{2}dw_{3},
\label{W1-pdf-permutation-identity}
\end{equation}
Therefore, considering Eq. (\ref{n-level-correlation-function-D-e})
one has:
\begin{eqnarray}
& & P_{3}(w_{1})=
\pi^{-3} \sum_{j_{1}=0}^{N-1} \sum_{j_{2}=0}^{N-1} \sum_{j_{3}=0}^{N-1}
\frac{1}{j_{1}!j_{2}!j_{3}!}I_{j_{1}j_{2}j_{3}}(w_{1}),
\label{W1-pdf-I-result} \\
& & 
I_{j_{1}j_{2}j_{3}}(w_{1})=
\int_{{\bf C}^{2}}
\exp(-\sum_{i=1}^{3} \vert (\Omega^{-1} w)_{i} \vert^{2})
\prod_{k=1}^{3}
\vert (\Omega^{-1} w)_{k} \vert^{2j_{k}}dw_{2}dw_{3}.
\label{W1-pdf-I-def}
\end{eqnarray}
One changes variables in Eq. (\ref{W1-pdf-I-def})
in following way: $V_{2}=2W_{2},V_{3}=-W_{3},$
and obtains:
\begin{eqnarray}
& & I_{j_{1}j_{2}j_{3}}(w_{1})=
\nonumber \\
& & 
= 2^{-2j_{2}}\int_{{\bf C}^{2}}
\exp(- \vert w_{1}+v_{2}+v_{3} \vert^{2}
- \frac{1}{4} \vert v_{2}\vert^{2}
- \vert v_{3} \vert^{2})
\vert w_{1}+v_{2}+v_{3} \vert^{2j_{1}}
\vert v_{2}\vert^{2j_{2}}
\vert v_{3} \vert^{2j_{3}} dv_{2}dv_{3}.
\label{W1-pdf-I-V2V3}
\end{eqnarray}
The above double integral can be calculated
by extending the exponent by additional
terms proportional to $\lambda_{i}$ parameters
and considering appropriate derivatives:
\begin{eqnarray}
& & I_{j_{1}j_{2}j_{3}}(w_{1})=
2^{-2j_{2}} 
\frac{\partial^{j_{1}+j_{2}+j_{3}}}
{\partial^{j_{1}} \lambda_{1} \partial^{j_{2}} \lambda_{2}
\partial^{j_{3}} \lambda_{3}}
F(w_{1},\lambda_{1},\lambda_{2},\lambda_{3}) \vert _{\lambda_{i}=0},
\label{W1-pdf-I-F} \\
& & F(w_{1},\lambda_{1},\lambda_{2},\lambda_{3})= 
\int_{{\bf C}^{2}}
\exp[G(w_{1},v_{2},v_{3},\lambda_{1},\lambda_{2},\lambda_{3})] dv_{2}dv_{3},
\label{W1-pdf-I-F-G} \\
& & G(w_{1},v_{2},v_{3},\lambda_{1},\lambda_{2},\lambda_{3})=
(\lambda_{1}-1) \vert w_{1}+v_{2}+v_{3} \vert^{2}
+(\lambda_{2}- \frac{1}{4}) \vert v_{2}\vert^{2}
+(\lambda_{3}- 1) \vert v_{3} \vert^{2}.
\label{W1-pdf-I-F-G-quadratic}
\end{eqnarray}
Finally, we derive $F(w_{1},\lambda_{1},\lambda_{2},\lambda_{3})$
by transformation of parametric quadratic form 
$G(w_{1},v_{2},v_{3},\lambda_{1},\lambda_{2},\lambda_{3})$
to canonical form and integrating over $v_{2}, v_{3}$:
\begin{equation}
F(w_{1},\lambda_{1},\lambda_{2},\lambda_{3})=
A(\lambda_{1},\lambda_{2},\lambda_{3})
\exp[-B(\lambda_{1},\lambda_{2},\lambda_{3}) \vert w_{1} \vert^{2}],
\label{W1-pdf-I-F-final}
\end{equation}
where
\begin{eqnarray}
& & A(\lambda_{1},\lambda_{2},\lambda_{3})=
\frac{(2\pi)^{2}}
{(\lambda_{1}+\lambda_{2}-\frac{5}{4}) 
\cdot (\lambda_{1}+\lambda_{3}-\frac{5}{4})-(\lambda_{1}-1)^{2}},
\label{W1-pdf-I-A} \\
& & B(\lambda_{1},\lambda_{2},\lambda_{3})=
(\lambda_{1}-1) \cdot \frac{2 \lambda_{1}-\lambda_{2}-\lambda_{3}+\frac{1}{2}}
{2 \lambda_{1}+\lambda_{2}+\lambda_{3}-\frac{9}{2}}.
\label{W1-pdf-I-B}
\end{eqnarray}
Hence, we obtained analytical formula
for distribution $P_{3}(w_{1})$ of second difference for N-dimensional
Ginibre ensemble combining 
Eqs. 
(\ref{W1-pdf-I-result}),
(\ref{W1-pdf-I-F}),
(\ref{W1-pdf-I-F-final}), 
(\ref{W1-pdf-I-A}), 
(\ref{W1-pdf-I-B}).
What is worth to be mentioned is that second difference's
distribution is a triple sum of zero-centred Gaussian distributions with
different widths. The distribution is again function of only
modulus $\vert w_{1} \vert$ of second difference and it has
global maximum at origin.

\section{Comparison}
\label{sect-comparison}
Finally, in order to compare our results
for second difference for Ginibre ensemble with previous ones
for Gaussian and Poissonian ensembles
we must rescale them  by division by appropriate magnitude. 
We consider such rescaled dimensionless
second differences in following way.
The mean values of second differences
either in real or in complex case are zero, hence we cannot
divide second differences by the mean values.
It follows that we divide real valued second differences
$\Delta^{2} E_{1}$ 
for GOE(3) ($\beta=1$),
for GUE(3) ($\beta=2$),
for GSE(3) ($\beta=4$), for PE ($\beta=0$),
by mean spacings $<S_{\beta}>$ calculated for those
ensembles, and we create new dimensionless
second differences:
\begin{equation}
C_{\beta} = \frac{\Delta^{2} E_{1}}{<S_{\beta}>},
\label{rescaled-second-diff-GOE-GUE-GSE-PE}
\end{equation}
respectively 
\cite{Duras 1996 PRE,Duras 1996 thesis,Duras 1999 Phys,Duras 1999 Nap}.
The probability density functions of $C_{\beta}$'s
were calculated for GOE, GUE, GSE, and PE
\cite{Duras 1996 PRE,Duras 1996 thesis,Duras 1999 Phys,Duras 1999 Nap}.
Since second difference for Ginibre ensemble is complex valued,
then we choose its real part $A_{1}$ for comparison with $C_{\beta}$'s.
One divides $A_{1}$ by analogue of $<S_{\beta}>$'s, which is
mean value $<R_{1}>$ 
of radius $R_{1}$ Eq. (\ref{Ginibre-polar-second-diff-result})
of $\Delta^{2} Z_{1}$:
\begin{equation}
<R_{1}> =\int_{0}^{\infty} r_{1} f_{R_{1}}(r_{1}) dr_{1}
=\frac{53}{64} \sqrt{6 \pi}.
\label{Ginibre-mean-spacing-R1-def} 
\end{equation} 
Hence:
\begin{equation}
X_{1}=\frac{A_{1}}{<R_{1}>},
\label{Ginibre-X1-def} 
\end{equation} 
is rescaled dimensionless $A_{1}$.
The probability distributions of $C_{\beta}$'s and of $X_{1}$
depend on the same real variable $x$ which is equal to
$\frac{\Delta^{2} e_{1}}{<S_{\beta}>}$,
$\frac{a_{1}}{<R_{1}>}$, respectively
($e_{1}$ is value of energy $E_{1}$).

The second differences' distributions
for Gaussian, Poisson, and Ginibre ensembles 
assume global maxima at origin and that they are unimodular.
Firstly, it extends
the theorem of level homogenisation
to Ginibre ensemble \cite{Duras 1996 PRE,Duras 1996 thesis,Duras 1999 Phys,Duras 1999 Nap}. 
We can formulate the following law: 
{\it Energy levels for Gaussian ensembles, for Poisson ensemble,
and for Ginibre ensemble tend to be homogeneously distributed.}
The second differences' distributions assume global maxima at origin
no matter whether second differences are real or complex.
From Coulomb gas' point of view it is easier
to be understood.
The unit charges behave in such a way that 
the vectors of relative positions of vectors
of relative positions of charges statistically tend to be zero.
It could be called stabilisation
of structure of system of electric charges. 
The above results can be extended 
to study of higher differences' distributions
for Ginibre ensemble.

\section{Acknowledgements}
\label{sect-acknowledgements}
It is my pleasure to deepestly thank Professor Jakub Zakrzewski
for formulating the problem.

\end{document}